\documentclass[aps,prl,twocolumn,amsmath,amssymb]{revtex4-1}
\usepackage{graphicx}
\usepackage{amsmath}
\usepackage{color}
\usepackage{dcolumn}
\usepackage{bm}
\usepackage{booktabs}
\usepackage{subfigure}
\usepackage{hyperref}
\hypersetup{
  colorlinks   = true,  
  urlcolor     = blue,  
  linkcolor    = blue,  
  citecolor    = red    
}

\begin{document}

\title{Stability of rubidium  molecules in the lowest triplet state}

\author{B.~J.~Verhaar}
\author{S.~J.~J.~M.~F.~Kokkelmans}
\affiliation{Eindhoven University of Technology, P. O. Box 513, 5600 MB 
Eindhoven, The Netherlands}

\date{\today}

\begin{abstract}
Experiments involving ultracold molecules require sufficiently long lifetimes, which can be very short for excited rovibrational states in the molecular potentials. For alkali atoms such as rubidium, a lowest rovibrational molecular state can both be found in the electronic singlet and triplet configurations. The molecular singlet ground state is absolutely stable. However, the lowest triplet state can decay to a deeper bound singlet molecule due to a radiative decay mechanism that involves the interatomic spin-orbit interaction. We investigate this mechanism, and find the lifetime of rubidium molecules in the lowest triplet rovibrational state to be about 21 minutes.
\end{abstract}

\maketitle

Stable ultracold molecules are of high experimental and theoretical interest 
\cite{Carr09}. In particular molecules with a permanent electric dipole moment 
offer the opportunity to explore many-body states \cite{Micheli07} that are 
impossible to reach with the isotropic nature of the short-range ultracold 
atomic interactions. One of the routes to create ultracold diatomic molecules is 
to associate them from ultracold atoms. Initially atoms are associated into 
weakly-bound Feshbach molecules by sweeping a magnetic field across resonance. 
Subsequently stimulated Raman adiabatic passage (STIRAP) is performed on these 
molecules to convert them to the lowest state of a particular potential \cite{Kokkelmans01}. This technique has proven to be very 
efficient, and in 2008 the first sample of diatomic KRb molecules in the 
rovibrational ground state was produced \cite{Ni08}. More recently, this also 
succeeded for RbCs \cite{Takekoshi14,Molony14}, which contrary to KRb is 
chemically stable under two-body collision processes \cite{Zuchowski10}. Also 
non-dipolar Rb$_2$ \cite{Lang08} and Cs$_2$ \cite{Danzl10} ground-state 
molecules have been created in this way.

The stability of these molecules is crucial for experiments, and therefore it is 
only natural to create the molecules in the absolute ground state. This is the 
rovibrational ground state of the electron spin singlet potential 
$X^1\Sigma_g$. The singlet potential is energetically very deep, 
and to reach its ground state via STIRAP, typically an additional laser system is required. However, the lowest spin triplet state is much less deep, and can be reached more easily with the laser set-up which is usually present for laser cooling and trapping purposes.

While (singlet) ground-state molecules are absolutely stable with respect to 
radiative decay, molecules in the lowest triplet state are not. However, the question is whether the radiative lifetime will be a practical limiting factor to current experiments. Recent experimental and theoretical work shows that a gas of singlet ground-state molecules has a very short reactive lifetime resulting from 3-body collisions \cite{Mayle13,Takekoshi14}. Also, triplet molecules have theoretically been shown to be unstable towards trimer formation \cite{Tomza13}. On the other hand, isolated Rb$_2$ molecules in the lowest triplet state, produced in an optical lattice \cite{Lang08}, are not sensitive to other types of decay apart from the radiative process, and may potentially have a much longer lifetime than the reactive lifetimes of both singlet and triplet molecules in their lowest rovibrational state.

In this paper, we investigate the lifetime of the lowest triplet Rb$_2$ state $a^3\Sigma^+_u$. While our approach is generic for all alkali atoms, rubidium is particularly interesting as it is currently the most-widely used species in ultracold quantum gas experiments. Our treatment applies to $^{85}$Rb$_2$, $^{87}$Rb$_2$, and to $^{85}$Rb$^{87}$Rb molecules.

Rb$_2$ molecules in the lowest triplet state are not absolutely stable, as the combined electron spin may form a lower energetic singlet configuration. An energy-conserving spin-flip, which would be a result of the magnetic dipole-dipole interaction \cite{Stoof88}, has a low probability due to the compact nature of a molecule in the lowest triplet state and the essential role of the nuclear magnetic moment both in this interaction (between a valence electron magnetic moment of one atom and the nuclear magnetic moment of the other), as well as in the subsequent nuclear spin M1 (magnetic dipole) decay \cite{Stoof88}. The lifetime associated with this mechanism would be roughly $10^5$ hours. 
\begin{figure}\begin{center}
 \includegraphics[width=\columnwidth]{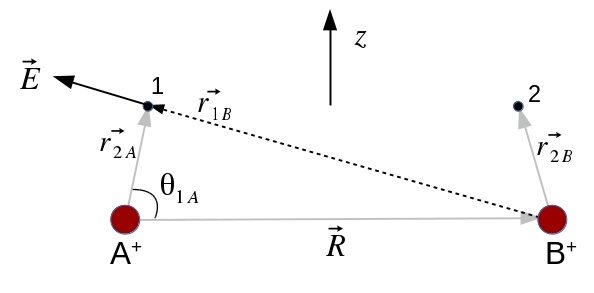}
 \caption{Geometry of Rb$_2$ molecule, considered to consist of two valence electrons 1,2, and two Rb$^+$ ions A$^+$,B$^+$. Electrons 1,2 and the atomic nuclei are initially spin-polarized in a direction $z$ (nuclear spins $i_A$ and $i_B$). Vector $\vec{E}(\vec{r}_{1B})$ is the Coulomb field at the position of electron 1 from the net charge of ion B$^+$ concentrated at nucleus B. The figure illustrates one of the four terms contributing to the interatomic spin-orbit interaction $V^{so}_{fi}$ (see Eq.~\ref{Vso1B}).}  
 \label{fig:mechanism}
\end{center}\end{figure}

However, there is a more probable spin-flip mechanism possible due to an inter-atomic part $V^{so}$ of the total spin-orbit interaction \cite{Mizushima,Itoh} that excites the two-atom system. The latter is considered to consist of two valence electrons 1,2 and two Rb$^+$ ions A,B with nuclear spins $i_A$,$i_B$. One of the terms contributing to $V^{so}$ in the situation of Fig.~\ref{fig:mechanism}, in which the electron-ion pair 1,B of the dimer is involved is \cite{Mizushima,Itoh}
\begin{equation}
 V^{so}(1,B) = \frac{e}{4m^2c^2} \left(\vec{E}(\vec{r}_{1B})
 \times \vec{p}_1\right).\vec{\sigma}_1 \equiv \frac{e\hbar}{4m^2c^2} \vec{V}_{1B}.\vec{\sigma}_1,
 \label{Vso1B}
\end{equation}
where we include the Thomas precession factor $\frac{1}{2}$ and define the shorthand $\vec{V}_{1B}$ for the spatial part of $V^{so}(1,B)$, i.e., $\vec{E}(\vec{r}_{1B}) \times \vec{p}_1$ (similarly for other electron-ion combinations). In Eq.~(\ref{Vso1B}) $\vec{E}(\vec{r}_{1B}) = e\hat{r}_{1B}/(4\pi\epsilon_0 r_{1B}^2)$ is the electric field operating on electron 1 due to the net charge of the other ion $B^+$ concentrated at its nucleus, $\hat{r}_{1B}$ is the unit vector $\vec{r}_{1B}/r_{1B}$, $\vec{\sigma}$ the Pauli spin vector, $\vec{p}_1$ the electronic momentum operator, $e$ the elementary charge, $c$ the velocity of light, and $m$ the electron mass. Rewritten in atomic units and including the 2A term we find:
\begin{equation}
 V^{so}(1,B) + V^{so}(2,A)= \frac{1}{4} \left(\frac{\lambdabar_c}{a_0}\right)^2 (\vec{V}_{1B}.\vec{\sigma}_1 + \vec{V}_{2A}.\vec{\sigma}_2),
 \label{Vso1B2A}
\end{equation}
with $\lambdabar_c$ the reduced electron Compton wavelength and $a_0$ the Bohr radius.

We study the excitation process at fixed values of the internuclear distance $R$ in an $R$ range (roughly equal to or larger than twice the Rb \textit{atomic} van der Waals radius $r_{vdW}=5.72a_0$ \cite{Mantina}), where it is reasonable to assume that electron 1 is in one atom and electron 2 in the other. We therefore define a pair of projection operators $\Pi$ on disjunct parts of 4-particle (two valence electrons and two ions) configuration space where either 1A,2B is the composition of the two atoms (projection operator $\Pi_{1A,2B}$) or 1B,2A ($\Pi_{1B,2A}$). We thus rewrite the above expression (\ref{Vso1B2A}) as
\begin{eqnarray}
 V^{so} &=& \frac{1}{4} \left(\frac{\lambdabar_c}{a_0}\right)^2 [\Pi_{1A,2B}(\vec{V}_{1B}.\vec{\sigma}_1 + \vec{V}_{2A}.\vec{\sigma}_2)
 \nonumber \\
 &+&\Pi_{1B,2A}(\vec{V}_{1A}.\vec{\sigma}_1 + \vec{V}_{2B}.\vec{\sigma}_2)]. 
 \label{Vso}
\end{eqnarray}
It is the $\vec{\sigma}_1 - \vec{\sigma}_2$ part $V^{so}_{fi}$ of $V^{so}$, proportional to the difference of the valence electron spins (antisymmetric in 1 and 2), that admixes a superposition of excited singlet two-particle electronic sp states into the initial fully spin-polarized initial dimer state $|\Psi_i \rangle$:
\begin{eqnarray}
V^{so}_{fi}&=&\frac{1}{8} \left( \frac{\lambdabar_c}{a_0} \right)^2  \left( \Pi_{1A,2B} (\vec{V}_{1B} - \vec{V}_{2A}) \right. \nonumber \\
&+& \left. \Pi_{1B,2A} (\vec{V}_{1A} - \vec{V}_{2B})\right) \cdot (\vec{\sigma}_1 - \vec{\sigma}_2). 
\label{Vsofi}
\end{eqnarray}

For the initial and final electronic states we take the $R$-dependent adiabatic potential-energy curves (PECs) and electronic transition dipole moments (TDMs) of the low-lying Rb dimer states from Ref.~\cite{Allouche12}. In that paper they are calculated both without and with (\textit{intra})-atomic spin-orbit coupling, leading to $\Lambda\Sigma$ states and $\Omega$ states, respectively. Inter-atomic spin-orbit coupling is not taken into account (see also Ref.~\cite{allouche92,Mueller84a,Mueller84b}). In our case the choice between the two types of states depends crucially on the range of inter-atomic distances where the initial state is concentrated. In Fig. 2 we present the PEC for the initial lowest triplet rovibrational state with quantum numbers $v_t,l,m_l=0,0,0$, together with the corresponding eigenfunction squared $[\phi_i(R)]^2$ with a Gaussian-like shape, normalized according to $\int_0^\infty [\phi_i(R)]^2 dR = 1$. It is concentrated in the interval from $R$ = 10 to 14 $a_0$, which contributes dominantly to the transition from $|\Psi_i\rangle$ to the intermediate states. In that range only $\Omega$-type $sp$ final states are relevant. In addition, the PECS for the $(1) 0^-_u$ and $(1) 1_u$ initial $\Omega$-states are virtually identical to those of the $a^3\Sigma^+_u$ state (for $R$ = 10 to 14 $a_0$ in 6 of 8 decimals). For that reason we will only make use of the $\Omega$-type TDMs in Ref.~\cite{Allouche12}. In addition, we only include them for $sp$ final states and omit the $sd$ states.

In order to connect to the TDMs of Ref. (\cite{Allouche12}) we now expand the factor $1/r_{1B}^3$ in the electric field $\vec{E}(\vec{r}_{1B}) = \vec{r}_{1B} /r_{1B}^3$ shown in Fig.~\ref{fig:mechanism} (similarly for other electric fields in the previous equations) in inverse powers of $R$, using the shorthands $\rho = r_1{_B}/R$, $p = -2 \cos\theta_{1A}~r_{1A}/R$, $q = (r_{1A}/R)^2$: 
\begin{eqnarray}
1/{\rho}^3&=& 1 - \frac{3}{2} (p+q) + \frac{15}{8} (p+q)^2 + ..... \nonumber \\
          &=& 1 + 3 \frac{z'_{1A}}{R} - \frac{3}{2} \left(\frac{r_{1A}}{R}\right)^2 
           + \frac{15}{2} \left(\frac{z'_{1A}}{R}\right)^2 + .....
\label{expansion}
\end{eqnarray}  
We use a nuclei-fixed right-handed coordinate system: the origin halfway the nuclei \cite{Allouche12}, an internuclear $z'$ axis, and a perpendicular pair of $x'$ and $y'$ axes with an arbitrary orientation around the $z'$ direction. Furthermore, we neglect cross-terms between the intra- and inter-atomic spin-orbit couplings (both weak), so that the electronic momentum in Eq.~(\ref{Vso1B}) can be expressed as a commutator of the Hamiltonian $H_{el}$ for the two valence electrons \cite{Allouche12} with the position vector $\vec{r}_{1A}$:  $\vec{p_1}=i[H_{el},\vec{r}_{1A}]$. As a consequence, we have
\begin{equation}
\vec{E}(\vec{r}_{1B}) \times \vec{p}_1 \propto [H_{el},\vec{r}_{2B} \times \vec{r}_{1A}] = [H_{el},-\vec{R} \times \vec{r}_{1A}], 
\end{equation}
i.e., only the component $\vec{r}_{1A\perp} = \vec{r}_{1\perp}$ perpendicular to  $\vec{R}$ and only even orders in the $1/\rho^3$ expansion survive.
The $0^{th}$ order contribution to $\vec{V}_{1B}$ can thus be dealt with in terms of TDMs and the $2^{nd}$ order term can be used to estimate the relative error, which turns out to be roughly $(r_{vdW}/R)^2 \approx 25\%$ in the relevant range $10 < R < 14a_0$ (see Fig.~\ref{fig:groundstate}). This conclusion is valid for other electron-ion combinations too.

The foregoing steps change the spatial operator multiplying $\Pi_{1A,2B}$ in Eq.~(\ref{Vsofi}) into  
$\vec{V}_{1B} - \vec{V}_{2A} = -\frac{i}{R^3} \vec{R} \times [H_{el},(\vec{r}_{1A} + \vec{r}_{2B})]$. A similar result is obtained for the operator multiplying $\Pi_{1B,2A}$. In total we obtain
\begin{eqnarray}
V^{so}_{fi} &=& \frac{1}{8R^2} \left( \frac{\lambdabar_c}{a_0} \right)^2 \hat{R} \times  
\left( \Pi_{1A,2B} [H_{el},\vec{r}_{1\perp} + \vec{r}_{2\perp}] \right. \nonumber \\
&+& \left. \Pi_{1B,2A} [H_{el},\vec{r}_{1\perp} + \vec{r}_{2\perp}]\right) \cdot (\vec{\sigma}_1 - \vec{\sigma}_2).
\label{Vsofia}
\end{eqnarray} 

The symmetry properties of this expression determine selection rules for the admixtures induced by $V^{so}_{fi}$. Splitting the electronic spatial part $\vec{D} = \vec{r}_{1\perp} + \vec{r}_{2\perp}$ of $V^{so}_{fi}$ into spherical components $q = \pm 1$ \cite{Brink}, $D_{\pm 1'} = \mp(D_{x'}\pm iD_{y'})/\sqrt{2}$, we find that they change parity $u$ into $g$ and in addition change the $z'$ component $M'_L$ of the total electronic by $\pm 1$, i.e., only a $1_g$ part is added to $0^-_u$ and only $0^+_g, 0^-_g, 2_g$ parts to $1_u$. We also find that each of the $D_{\pm 1'}$ terms changes $\sigma_v$ reflection parity \cite{Mizushima} from + to - and vice versa. The foregoing implies that we can use the $\Delta \Omega = \pm 1$ transition dipole moments (TDMs) for E1 transitions published by Allouche et al. \cite{Allouche12} to calculate the $(\vec{r}_1 + \vec{r}_2)_{\pm 1'}$ spatial matrix elements, in combination with the spin matrix elements $\langle(S,M'_S)_f = 0,0|(\vec{\sigma}_1-\vec{\sigma}_2)_{\mp 1'}|(S,M'_S)_i = 1,\pm 1 \rangle$. The equality $q=M'_S$ illustrates angular momentum conservation along the $z'$ symmetry axis: spin angular momentum is transferred to orbital angular momentum. 

A necessary following step is to impose Kronig symmetry\cite{Kronig}: we require invariance of Hamilton operator and wave functions under the combination of a rotation of the nuclei over $\pi$ around the $y'$ axis (leaving the electrons alone) and space inversion of the electronic position coordinates with respect to the origin. Both $H_{el}$ and each of the $\Pi_{1A,2B}$ and $\Pi_{1B,2A}$ terms in Eq.~(\ref{Vsofia}) obey this invariance. As a consequence transitions induced by $V^{so}_{fi}$ take place between states with equal Kronig symmetry only. To find the Kronig symmetric and antisymmetric $(1)0^-_u$ and $(1)1_u$ states, we make use of our earlier conclusion that in the $R$ interval of interest these $\Omega$ states are very close to $^3\Sigma^+_u$ states, i.e., $\Lambda\Sigma$ states with $\Lambda = 0$ and $\Sigma = 0, \pm 1$. We therefore equate them to the corresponding $\Omega$ states. In Herzberg's notation \cite{Herzberg,Mizushima}
\begin{eqnarray}
|(1)0^-_u\rangle &=& |(1)^3\Sigma^+_u,\Omega=\Sigma=0\rangle \nonumber \\
                 &=& |c,(1)^3\Sigma^+,S=1,\Omega=0\rangle
\label{c0u}
\end{eqnarray}
is a Kronig-symmetric state by itself, as indicated by the symbol '$c$', whereas $|(1)^3\Sigma^+_u,\Omega=1\rangle$ falls apart as a normalized sum and difference of the following Kronig-symmetric $c$ and antisymmetric $d$ parts:
\begin{eqnarray}
|^c_d,(1)1_u\rangle &=& [|\Sigma^+,S=1,\Omega=1,\Sigma=+1\rangle  \nonumber \\
 &\pm& |\Sigma^+,S=1,\Omega=1,\Sigma=-1\rangle]/\sqrt{2}.
\label{cd1u}
\end{eqnarray}
We conclude that the above-mentioned selection rules have to be further specified: allowed transitions are
$(c,0^-_u)$  $\rightarrow$  $(c,1_g)$, $(c,1_u)$  $\rightarrow$  $(c,0^+_g)$ or $(c,2_g)$, and $(d,1_u)$   $\rightarrow$  $(d,0^-_g)$ or $(d,2_g)$.

Each of the above $c$ and $d$ states (\ref{c0u}) and (\ref{cd1u}), multiplied by $|S,M_S = 1,+1\rangle$, is present initially with probability $1/3$ before the excitation by $V^{so}_{fi}$ and has the total form
\begin{eqnarray}
 |\Psi_{i,n_i=1}(\vec{R})\rangle &=& \psi_i(n_i=1; R) Y_{0,0}(\theta_R,\phi_R) \nonumber \\
 &\otimes& |S,M_S=1,+1\rangle,
 \label{Psii}
\end{eqnarray}
with $\psi_i(n_i; R)$ standing for the $R$-dependent spatial part of the three initial symmetry types distinguished by $i$, and $n_i$ being the serial number of the corresponding $H_{el}$ eigenvalues counting from below. A similar notation is used for the final states. Furthermore, $Y$ is a spherical harmonic depending on spherical angles that specify the direction of $\vec{R}$. We leave out the nuclear spin state $|I,M_I=i_A+i_B,i_A+i_B\rangle$, which is not affected in the excitations and subsequent decay processes. The quantum numbers $M_S$ and $M_I$ are projections of the total electron spin $S$ and total nuclear spin $I$ on the $z$-axis along which the polarized initial state has been prepared. As a further step we expand the initial electronic spin state $|1,+1\rangle$ in states $|1,M'_S\rangle$ quantized along the direction of the $z'$ axis:
\begin{equation}
|S,M_S=1,+1\rangle = \sum_{M'_S} D^{1^{*}}_{1,M'_S}(\phi_R,\theta_R,\chi_R)~|1,M'_S\rangle,
\label{spinstate}
\end{equation}
where $M'_S = 0,\pm1$. Furthermore, $\chi_R$ is an angle around $\hat{z}'$ further specifying the directions of the $x'$ and $y'$ axes and completing the angles $\theta_R,\phi_R$ in Eq.~(\ref{Psii}) to a set of Euler angles  \cite{Brink}. Each of the final states has the form 
\begin{eqnarray}
 \Psi_{f,n_f}(\vec{R}) &=& \psi_f(n_f; R) \sqrt{\left[\frac{3}{8\pi^2}\right]} D^{1^{*}}_{1,M'_S}(\phi_R,\theta_R,\chi_R) \nonumber \\
 &\otimes& |S,M_S=0,0\rangle.
 \label{Psif}
\end{eqnarray}
The spatial wave functions $\psi_i(n_i; R)$ and $\psi_f(n_f; R)$ form an orthonormal set of eigenfunctions of the Hamiltonian $H_{el}$ for the valence electrons.

For given $R$ and $M'_S$ we now consider the transition amplitude induced by  $V^{so}_{fi}$ between an initial state $\Psi_{i,n_i}(R)$ and a final state $\Psi_{f,n_f}(R)$:
\begin{eqnarray}
A_{fn_f,in_i}(M'_{S},R)&=& a_{f,i} b_{f,i} \\
                       &.& \sqrt{\left[\frac{3}{8\pi^2}\right]}
                       \frac{D^{1^{*}}_{1,M'_S}(\phi_R,\theta_R,\chi_R)}{E_{f,n_f}-E_i,n_i}. \nonumber
\label{Ampl}
\end{eqnarray}
The factors $a_{f,i}$ and $b_{f,i}$ are spatial and spin matrix elements: $a_{f,i}=\langle\psi_f(n_f; R)|[H_{el},(\vec{r}_1+\vec{r}_2)_q]|\psi_i(n_i; R)\rangle$ and $b_{f,i}=\langle 0,0|(\vec{\sigma}_1-\vec{\sigma}_2)_{-q}|1,M'_{S}\rangle$, in which $q=M'_S=\Omega_f-\Omega_i=\pm 1$, depending on the symmetries $f$ and $i$. In $a_{f,i}$, letting $H_{el}$ operate to the left in one term of the commutator and to the right in the other, we find that this results in a factor $E_{f,n_f}-E_{i,n_i}$ that cancels out the denominator in Eq.~(\ref{Ampl}). The energies $E$ are $R$-dependent adiabatic potential-energy values (eigenvalues of $H_{el}$, see PECS in Ref.~\cite{Allouche12}) and $|0,0\rangle, |1,M'_{S}\rangle$ are valence electron spin states. Furthermore, to make angular momentum conservation more transparent we have converted the Cartesian inner product $(\vec{r}_{1\perp}+\vec{r}_{2\perp})\cdot(\vec{\sigma}_1 - \vec{\sigma}_2)$ in Eq.~(\ref{Vsofia}) to products of spherical components in the body-fixed system
$\sum_{q=-1}^{+1} \delta_{|q|,1}(-1)^q (\vec{r}_{1,q}+\vec{r}_{2,q}) (\vec{\sigma}_{1,-q}-\vec{\sigma}_{2,-q})$.: for a non-vanishing result $q$ has to be equal to $M'_S$.

For comparision reasons, we discuss here an alternative approach that we investigated. Here we started from the full $V^{so}$ of Eq.~({\ref{Vso}) operating on the initial state (\ref{Psii}), in which we took the radial wave function of the Rb 5s valence electron from \cite{Szasz67}, leading to a sum of $\Lambda S \Sigma \pi_e= 1 0 0 g$ and $1 1 -1 g$ parts. Subsequently, we followed \cite{Movre77} in describing the E1 decay. The sum of the intra-atomic spin-orbit couplings and the inter-atomic electric dipole-dipole interaction $V^{dd}$ was diagonalized in the 18 dimensional space, leading to a set of 18 $R$-dependent eigenstates in the 5s5p space considered in Ref.~\cite{Allouche12}, comparable to the final states $\psi_f(n_f; R)$ above. The main shortcoming of this approach is the role of the interaction $V^{dd}$ in the radial range 10 - 14$a_0$, where it is a bad approximation. Due to the strong repulsion in some of the final states and a strong attraction in the remaining ones the radial wave functions has small values. This leads to a lifetime of the lowest-energy Rb$_2$ triplet state of about 25 hours, which is three orders of magnitude larger than wat we calculate below.

Continuing the present treatment based on Ref.~\cite{Allouche12}, we take the absolute square of the above amplitude $A$ in Eq.~(\ref{Ampl}), integrate over the Euler angles, and use the normalization of the $D$-function \cite{Brink}. We thus find
\begin{widetext}
\begin{equation}
\int |A_{fn_f,in_i}(M'_{S},R)|^2 \sin({\theta}_R) d\phi_R d\theta_R d\chi_R =
 \frac{1}{4R^4}.\frac{8\pi^2}{3}.\left( \frac{\lambdabar_c}{a_0} \right)^4 \left[{\rm TDM}(f,n_f;i,n_i;R)\right]^2.
\end{equation}
\end{widetext}
In our notation the $TDM$ matrix element in this equation is given by 
\begin{equation}
{\rm TDM}(fn_f,in_i;R) = |\langle \psi_f(n_f;R)| [\vec{r}_{1q} + \vec{r}_{2q}]|\psi_i(n_i; R)\rangle|, 
\end{equation}
with the spherical component $q = \Omega_f - \Omega_i$, the change of the $\Omega$-values from initial to final states. The actual TDM values \cite{Allouche12} show which final states are primarily excited starting from the three initial states. For larger $R$ (beyond $40 a_0$) only the three lowest-energy $c,1_g$ states are significantly excited from $0^-_u$, as well as the (2)-(3) $0^+_g$ and (1) $2_g$ state from $1_u$ (for notation see \cite{Allouche12}; (1) $0^+_g$ = absolute singlet ground state). More relevant for our purpose, in the above-mentioned interval $R$ = 10 to 14 $a_0$ the $V^{so}_{fi}$ strength is distributed among transitions from $0^-_u$ to the sp states (1)-(2) $1_g$, as well as from $1_u$ to (2)-(3) $0^+_g$, to (1)-(2) $0^-_g$, and to (1)$2_g$. All of these $sp$ states undergo E1 decay back to the $ss$ states.

Averaging over the three initial $i,n_i$ states (\ref{c0u}), (\ref{cd1u}), summing over the final $f,n_f$ states, and multiplying by the average $\gamma = 3.60 \times 10^7 s^{-1}$ of the Rb atomic first excited $^2P_{1/2}$ and $^2P_{3/2}$ spontaneous decay rates, we find our estimate of the local decay rate $\Gamma(R)$ of the lowest triplet Rb$_2$ state,
\begin{eqnarray}
\Gamma(R) &=& \gamma \frac{1}{4R^4}.\frac{8\pi^2}{3}.\left( \frac{\lambdabar_c}{a_0} \right)^4 \sum_{in_i} \sum_{fn_f} \left( \frac{1}{3}\delta_{i,0^-_u} \right. \nonumber \\
&+& \left. \frac{2}{3}\delta_{i,1_u} \right) \left[{\rm TDM}(fn_f,in_i;R)\right]^2, 
\label{gammar}
\end{eqnarray}
displayed in Fig.~\ref{fig:groundstate}, and the total decay rate
\begin{equation}
\Gamma = \int_0^\infty \langle \phi_i(R)|\Gamma(R)|\phi_i(R) \rangle dR.
\label{gamma}
\end{equation}
Our result for the total :decay rate is $0.78 \times 10^{-3}s^{-1}$, corresponding to a lifetime of about 1280 s = 21 min.  

\begin{figure}\begin{center}
  \includegraphics[width=\columnwidth]{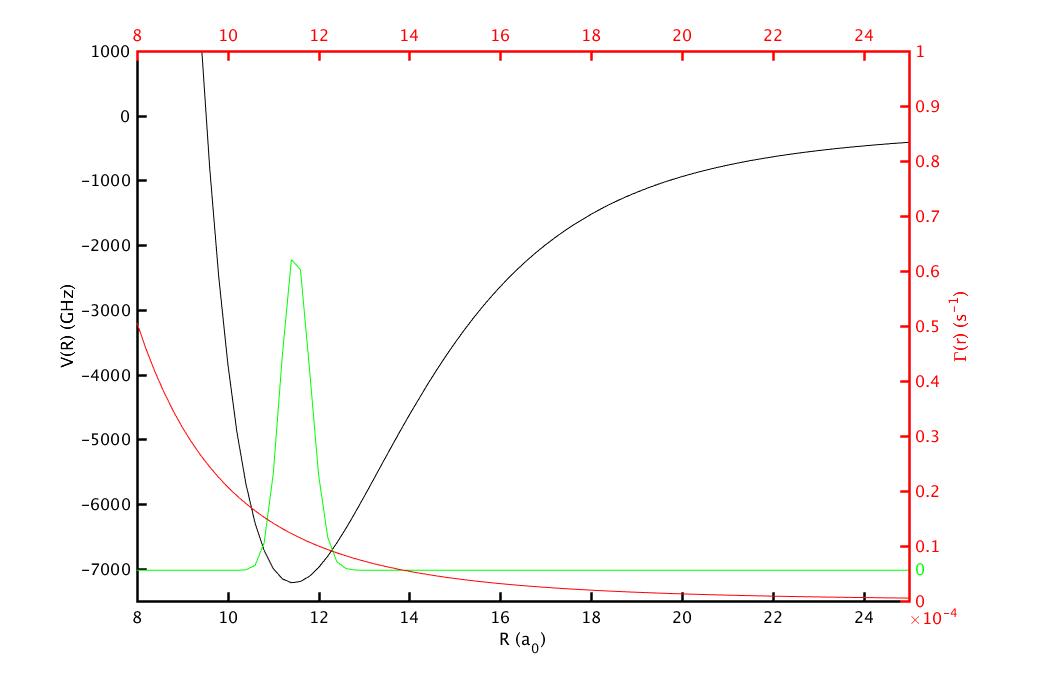}
  \caption{Potential for lowest Rb$_2 0^-_u$ and $1_u$ states, where the rovibrational ground state is located at -7.026 GHz. The corresponding squared wave function $[\phi_i(R)]^2$ is indicated (green). 
  Also indicated is the local decay rate $\Gamma(R)$ (red).} 
  \label{fig:groundstate}
\end{center}\end{figure}

We conclude that isolated rubidium molecules in the lowest-energy triplet state have a finite lifetime. This is due to a radiative mechanism involving an inter-atomic spin-orbit interaction and inducing decay to the singlet state. The lifetime is about 21 min., which is much longer than typical experimental time scales needed to study these ultracold molecules, created in an optical lattice starting from two atoms on each lattice site. Other decay mechanisms resulting in a transition to deeper-bound singlet states involve collisions, for instance with other triplet ground-state molecules. Future experiments should be able to investigate this mechanism, and shed more light on the collisional lifetime.

\begin{acknowledgments}
We gratefully acknowledge discussions with William Stwalley and Andrei Derevianko. This research was supported in part by the National Science Foundation under Grant No. NSF PHY11-25915.
\end{acknowledgments}

\bibliographystyle{apsrev}
\bibliography{rbtripletmolecules}

\end{document}